\documentclass[11pt]{article}
\usepackage{graphicx}
\newcommand{\be}{\begin{equation}}
\newcommand{\ee}{\end{equation}}
\newcommand{\ber}{\begin{eqnarray}}
\newcommand{\eer}{\end{eqnarray}}

\title{Computational models of long term plasticity and memory \thanks{draft of an article that is being considered for publication by Oxford University Press in the forthcoming book "Oxford Research Encyclopedia of Neuroscience", Editor S. Murray Sherman, due for publication in 2017.}}
\author{Stefano Fusi \\ \small Center for Theoretical Neuroscience, College of Physicians
	and Surgeons, Columbia University\\ \small Mortimer B. Zuckerman Mind Brain Behavior Institute, Columbia University\\ \small Kavli Institute for Brain Sciences, Columbia University }

\begin{document}
	
	\maketitle


\section{Introduction}

Memory is often defined as the mental capacity of retaining information about facts, events, procedures and more generally about any type of previous experience. Memories are remembered as long as they influence our thoughts, feelings, and behavior at the present time. Memory is also one of the fundamental components of learning, our ability to acquire any type of knowledge or skills. 

In the brain it is not easy to identify the physical substrate of memory. Basically, any long-lasting alteration of a biochemical process can be considered a form of memory, although some of these alterations last only a few milliseconds, and most of them, if taken individually, cannot influence our behavior. However, if we want to understand memory, we need to keep in mind that memory is not a unitary phenomenon, and it certainly involves several distinct mechanisms that operate at different spatial and temporal levels. 

One of the goals of theoretical neuroscience is to try to understand how these processes are orchestrated to store memories rapidly and preserve them over a lifetime. Theorists have mostly focused on synaptic plasticity, as it is one of the most studied memory mechanisms in experimental neuroscience and it is known to be highly effective in training artificial neural networks to perform real world tasks. Some of the synaptic plasticity models are purely phenomenological and they have proved to be important for describing quantitatively the complex and rich observations in experiments on synaptic plasticity. Some other models have been designed to solve computational problems, like pattern classification, or simply to maximize the memory capacity in standard benchmarks. Finally, there are models are inspired by biology, but then find an application to a computational problem, or vice versa, there are models that solve complex computational problems that then are discovered to be biologically plausible. In this article I will review some of these models and I will try to identify computational principles that underlie memory storage and preservation (see also \cite{fiete16} for a recent review that focuses on similar issues).

\section{Long term synaptic plasticity}

\subsection{Abstract learning rules and synaptic plasticity}

Artificial neural networks are typically trained by changing the parameters that represent the neuronal activation thresholds and the synaptic weights that connect pairs of neurons. The algorithms used to train them can be divided into three main groups (see e.g. a classic textbook like \cite{hkp91}) 1) networks that are able to create representations of the statistics of the world in an
autonomous way (unsupervised learning) 2) networks that can learn to perform a particular task when instructed by a teacher (supervised learning) 3) networks that can learn by a trial and error procedure (reinforcement
learning). These categories can have a different meaning and different nomenclature depending on the community (machine learning or theoretical neuroscience). For all these algorithms, memory is a fundamental component which typically resides in the pattern of synaptic weights and in the activation thresholds. Every time these parameters are modified, the memory is updated.

\subsubsection{The perceptron}

Rosenblatt \cite{r58, Rosenblatt62} introduced in the 60's one of the fundamental algorithms for training neural networks. He studied in detail what is probably the simplest feed-forward neural 'network', and the fundamental building block of more complex networks. He called it the perceptron. The perceptron is just one output neuron that is connected to $N$ input neurons. For a given input pattern $x^\mu$ ($x^\mu$ is a vector, and its components $x^\mu_i$s are the activation states of specific neurons), the total current into the output neuron is a weighted sum of the inputs:

$$I^\mu=\sum_{i=1}^N w_{i} x_i^\mu$$

The output neuron can be either active or inactive. It is activated by the input only when $I$ is above an activation threshold $\theta$. 

The learning algorithm is supervised and it can be used to train the perceptron to classify input patterns into two distinct categories. During learning, the synaptic weights and the activation threshold are tuned so that the output neuron responds to each input as prescribed by the supervisor. For example, consider the classification problem in which the inputs represent images of handwritten digits and the perceptron has to decide whether a digit is odd or even. During training the perceptron is shown a large number of samples of odd and even digits, and the output neuron is set by the supervisor to the activation state corresponding to the class to which the input belongs (e.g. the neuron is activated when the digit is odd, inactivated when it is even). 

The learning procedure ensures that after learning the perceptron responds to an input as prescribed by the supervisor, even in its absence. The input can be one of the samples used for training, or a new sample from a test set. In the second case the perceptron is required to generalize and classify correctly also the new inputs (e.g. a new handwritten digit). 

The proper weights and the threshold are found using an iterative algorithm: for each input pattern, there is a desired output provided by the supervisor, which is $y^\mu$ ($y^\mu=-1$ for input patterns that should inactivate the output neuron and $y^\mu=1$ for input patterns that should activate the output neuron), and each synapse $w_{i}$, connecting input neuron $i$ to the output is updated as follows:

\begin{equation}
w_{i} \to w_{i} + \alpha x^\mu_i y^\mu
\label{perceptron}
\end{equation}

where $\alpha$ is a constant that represents the learning rate.
The threshold $\theta$ for the activation of the output neuron is modified in a similar way:

$$\theta \to \theta - \alpha y^\mu$$

The synapses are not modified if the output neuron already responds as desired. In other words, the synapse is updated only if the total synaptic current $I^\mu$ is below the activation threshold $\theta$ when the desired output $y^\mu= + 1$ (and analogously when $I^\mu>\theta$ and $y^\mu=-1$). The synapses are updated for all input patterns, repeatedly, until all the conditions on the output are satisfied. These synaptic updates are a simple form of synaptic plasticity.

The importance of the perceptron algorithm resides in the fact that it can be proved \cite{block1962} to converge if the patterns are linearly separable (i.e. if there exists a $w_{i}$ and a threshold $\theta$ such that $I^\mu>\theta$ for all $\mu$ such that $y^\mu=1$ and $I^\mu<\theta$ for all $\mu$ such that $y^\mu=-1$). In other words, if a solution to the classification problem exists, the algorithm is guaranteed to find one in a finite number of iterations. The convergence proof is probably one of the earliest elegant results of computational neuroscience.

\subsubsection{Hebb's principle}

The perceptron algorithm is also considered one of the early implementations of Hebb's principle \cite{hebb49}. The principle reflects an important intuition of Donald Hebb about a basic mechanism for synaptic plasticity. It states:

{\sl ``When an axon of cell A is near enough to excite a cell B and
	repeatedly or persistently takes part in firing it, some growth
	process or metabolic change takes place in one or both cells such
	that A's efficiency, as one of the cells firing B, is increased.''}

The efficiency he refers to can be interpreted as the synaptic efficacy, or the weight $w_i$ that we defined above. The product of the activities of pre and post-synaptic neurons that appear in the synaptic update equation Eq.\ref{perceptron} is often considered an expression of the Hebbian principle: when the input and the output neuron (pre and post-synaptic, respectively) are simultaneously active, the synapse is potentiated. In the case of the perceptron, the output neuron is activated by the supervisor during training and it reflects the desired activity. 

\subsubsection{Extensions of the perceptron algorithm}

Synaptic models that are biologically plausible implementations of the perceptron algorithms have been proposed in the last decades (see e.g.\cite{bsf05, legenstein2005}). In these models the neuronal activity is often expressed as the mean firing rate, but there are models that consider the timing of individual spikes. An interesting class of spike driven synaptic models solves the computational problem of how to train neurons to classify spatio-temporal patterns of spikes. For example, the tempotron algorithm, introduced by G\"utig and Sompolinsky in 2006 \cite{gs06}, can be used to train a spiking neuron to respond to a specific class of input patterns by firing at least once during a given time interval. The neuron remains silent in response to all the other input patterns. A recent extension of the model can be trained to detect a particular clue (basically a specific spatio-temporal pattern) simply by training it to fire in proportion to the clue's number of occurrences during a particular time interval \cite{g16}.

Besides the perceptron, there are other learning algorithms that are based on similar principles and often the synaptic weights are modified on the basis of the covariance between the pre and post synaptic activity (see e.g. \cite{sejnowski77,h82}). Many of these algorithms can be derived from first principles, for example by minimizing the error of the output. 

Error minimization is also the basic principle of a broad class of learning algorithms that can train artificial neural networks that are significantly more complex than the perceptron. For example feed-forward networks with multiple layers (deep) can be trained by computing the error at the output and backpropagating it to all the synapses of the network. This algorithm, called backpropagation \cite{rumelhart1986learning}, has recently revolutionalized machine vision, and is extremely popular in artificial intelligence \cite{LeCun2015}. Although it is difficult to imagine how backpropagation can be implemented in a biological system, several groups are working on versions of the algorithm which are more biological plausible (see e.g.\cite{Lillicrap2016, Scellier2016}). In the future these models will certainly play an important role in understanding synaptic dynamics and learning in the biological brain.

\subsection{Phenomenological synaptic models}

The synaptic models mentioned in the previous section were designed to implement a specific learning algorithm. However, there are also several models that were initially conceived to describe the rich phenomenology observed in experiments on synaptic plasticity. 


A popular class of models was inspired by the experimental observations that long term synaptic modifications depend on the precise timing of pre- and post-synaptic spikes\cite{ls83,mlfs97,bp98,stn01}. These models are usually described using the acronym STDP (Spike Timing Dependent Plasticity), introduced by Larry Abbott and colleagues to designate a specific model\cite{sma00}. In the simplest version of STDP, when the pre-synaptic spike precedes a post-synaptic spike within a time window of 10-20 ms, the synapse is potentiated and for the reverse order of occurrence of pre and post-synaptic spikes, the synapse is depressed. Phenomenological models can describe accurately many other aspects of the experimental observations \cite{Senn2001,Pfister2006,Morrison2008,Clopath2010}.
Several theoretical studies describe the dynamics of neural circuits whose synapses are continuously updated using STDP\cite{sma00,sb01,ba10,ba13}. The role of STDP in learning has also been investigated in many computational studies (see e.g. \cite{gkhw96,lnm05,gs06,Izhikevich2007a,Legenstein2008,Nessler2013,Pecevski2016}). One of the theoretical works \cite{gkhw96} preceded the experimental papers on STDP and hence predicted the STDP observations.

Although the computational principles behind STDP are probably general and important, experiments on synaptic plasticity show that STDP is only one of the aspects of the mechanisms for the induction of long term changes (see e.g. \cite{sww10}). The direction and the extent of a synaptic modification depend on various types of 'activity' of the pre and post-synaptic neurons (not only on spike timing), on the location of the synapses on the dendrite \cite{Sjoestroem2006}, on neuromodulators \cite{squirekandel,SjostromHausser2008}, on the timescales that are considered in the experiment (e.g. homeostatic plasticity occurs on timescales that are significantly longer than those of long term synaptic plasticity\cite{Turrigiano2004}) and, most importantly, on the history of previous synaptic modifications, a phenomenon called meta-plasticity \cite{Abraham2008} (see also below). 

\begin{figure}[htb]
	\begin{center}
		\includegraphics[width=13cm]{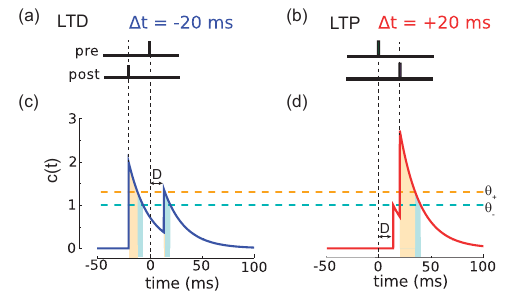}
		\caption{\small Calcium based synaptic model proposed in \cite{gb11}: long term depression (LTD, left) and long term potentiation (LTP, right). The spike timing dependent plasticity (STDP) protocol for inducing long term modifications are illustrated at the top in panels {\bf (a)} and {\bf (b)}: when the post-synaptic spike precedes the pre-synaptic spike, LTD is induced. The synapse is modified in the opposite direction (LTP) when the pre-synaptic spike precedes the post-synaptic action potential. This dependence of the sign of the synaptic modification on the relative timing of pre and post synaptic spikes can be obtained by introducing an internal variable that represents the post-synaptic calcium concentration $c(t)$. The weight increases when $c$ is above a threshold $\theta_+$ (orange in the figure) and it decreases when $c$ is between $\theta_-$ (cyan) and $\theta_+$. The calcium variable jumps to a higher value every time a pre-synaptic or a post-synaptic spike arrives. It does so instantaneously for post-synaptic spikes, and with a delay $D \sim 10 ms$ for pre-synaptic spikes. It then decays exponentially with a certain time constant, of the order of a few ms. 
		In panels {\bf (c)} and {\bf (d)}, $c(t)$ is plotted as a function of time in the LTD and LTP protocols based on STDP. When the post-synaptic spike precedes the pre-synaptic one, the calcium variable $c(t)$ spends more time between the orange and the cyan lines than above the orange line, eventually inducing long term depression (panel {\bf (c)}). When the pre-synaptic spike precedes the post-synaptic action potential (panel {\bf (d)}), most of the time the calcium variable is above the orange line and LTP is induced. Figure
			adapted from \cite{gb11}. }
	\end{center}
	\label{nicolasfig}
\end{figure}

A recent synaptic model can reproduce part of the rich STDP phenomenology observed in multiple experiments using a surprisingly small number of dynamical variables which include calcium concentration \cite{gb11} (see Fig. \ref{nicolasfig} for a description of the model and an explanation of how it can reproduce STDP).

\section{Memory}

The synaptic plasticity models described in the previous section can predict to some extent the sign of the synaptic modification. However, these models focus on one of the early phases of synaptic plasticity. The consolidation and the maintenance of synaptic modifications require a complex molecular machinery that typically involves cascades of biochemical processes that operate on different timescales. Typically the process of induction of long term synaptic modifications starts with an alteration of some of the molecules that are locally present at the synapse. For example, in Fig. \ref{nicolasfig} we described a model in which calcium concentration increases at the synapse at the arrival of  either the pre-synaptic spike or the post-synaptic back propagating action potential. The entry of calcium then induces an alteration of the state of some relatively complex molecules that are locally present, like the calcium/calmodulin dependent protein kinase II (CaMKII). In the case of CaMKII, each molecule can be either phosphorylated (activated) or unphosphorylated (inactived) and hence it can then be considered as a simple memory switch. The population of the few (5-30) CaMKII molecules that are present and immobilized in a dendritic spine are a basic example of a molecular memory. Their state modulates the strength of the synaptic connection.

One of the problems of molecular memories is related to their stability. Because of molecular turnover (see e.g.\cite{crick1984}), the memory molecules are gradually destroyed and replaced by newly synthesized ones. To preserve the stored memories, the state of the old molecules should be copied to the incoming naive ones. If not, the memory lifetime is limited by the lifetime of the molecule, which, in the case of CaMKII is of the order of 30 hours. Other molecules can last longer, but none of them can survive a lifetime. One possible explanation for long memory lifetimes is bistability, which was already proposed by Francis Crick \cite{crick1984}. For example, in the case of CaMKII one can imagine that the populations of all molecules has two stable points: one in which none of the molecules is active, and another one in which a large proportion is active. The dynamics of models describing this form of bistability have been studied in detail \cite{lisman1985,lismanzhabotinsky2001,miller2005}.
When a new inactivated protein comes in, it remains unaltered if the majority of the existing CaMKII molecules are inactivated, and it is activated if they are in the active state. As a consequence, the new molecules can acquire the state of the existing ones, preserving the the molecular memory at the level of the population of CaMKII molecules.

CaMKII is only one of the numerous molecules that are involved in synaptic plasticity: more than 1000 different proteins have been identified in the post synaptic proteom of mammalian brain excitatory synapses (see e.g. \cite{EmesGrant2012}). Interestingly, less than 10\% of these proteins are neurotransmitter
receptors, which suggests that the majority of proteins are not directly involved in electrophysiological functions and instead have signaling and regulatory roles. CaMKII is known to be important in the early phases of the induction of long term synaptic potentiation (E-LTP). The molecules involved in E-LTP activate a cascade of biochemical processes that eventually regulate gene transcription and protein synthesis, leading to permanent changes in the morphology of the synaptic connections or to persistent molecular mechanisms that are known to underlie late long term potentiation (L-LTP) maintenance. The foundational work on these cascades of biochemical processes of Eric Kandel and colleagues is summarized in \cite{squirekandel}.

To understand the computational role of these highly organized protein networks, it is necessary to review more than 30 years of theoretical studies. The next sections summarize some of the important results of these studies that show that biological complexity plays a fundamental role in maximizing memory capacity.

\subsection{Memory models and synaptic plasticity}

For many years, research on the synaptic basis of memory focused on the long-term potentiation of synapses which, at least by the modeling community, was represented as a simple switch-like change in synaptic state. Memory models studied in the 1980’s (i.e. \cite{h82}) suggested that networks of neurons connected by such switch-like synapses could maintain huge numbers of memories virtually indefinitely. Although it becomes progressively more difficult to retrieve memories in these models as time passes and additional memories are stored, the memory traces of old experiences never fade away completely (see e.g. \cite{a89,hkp91}. Memory capacity, which was computed to be proportional to network size, was only limited by interference from multiple stored memories, which can hamper memory retrieval. This work made it appear that extensive memory performance could arise from a relatively simple mechanism of synaptic plasticity. However, it was already clear from the experimental works summarized above that synaptic plasticity is anything but simple. If, as the theoretical work suggested, this complexity is not needed for memory storage, what is there for? 
 
The key to answering this question arose from work done at the beginning of the 90’s. This work arose from a project led by D. Amit aimed at implementing an associative neural network in an electronic chip using the physics of transistors to emulate neurons and synapses, as originally proposed by Carver Mead \cite{mead}. The main problem encountered in this project was related to memory. The problem was not how to preserve the states of synapses over long times, but how to prevent memories from being overwritten by other memories. Memories were overwritten by other memories so rapidly that it was practically impossible for the neural network to store any information. Subsequent theoretical analysis of this problem \cite{af92,af94,f02,fa07} showed that what had appeared to be a simple approximation made in the theoretical calculations of the 80s was actually a fatal flaw. The unfortunate approximation was ignoring the limits on synaptic strength imposed on any real physical or biological device. When these limits are included, the memory capacity grows only logarithmically rather than linearly with network size, and the models could no longer account for actual biological memory performance. 

\begin{figure}
\centerline{\includegraphics[width=3.5in]{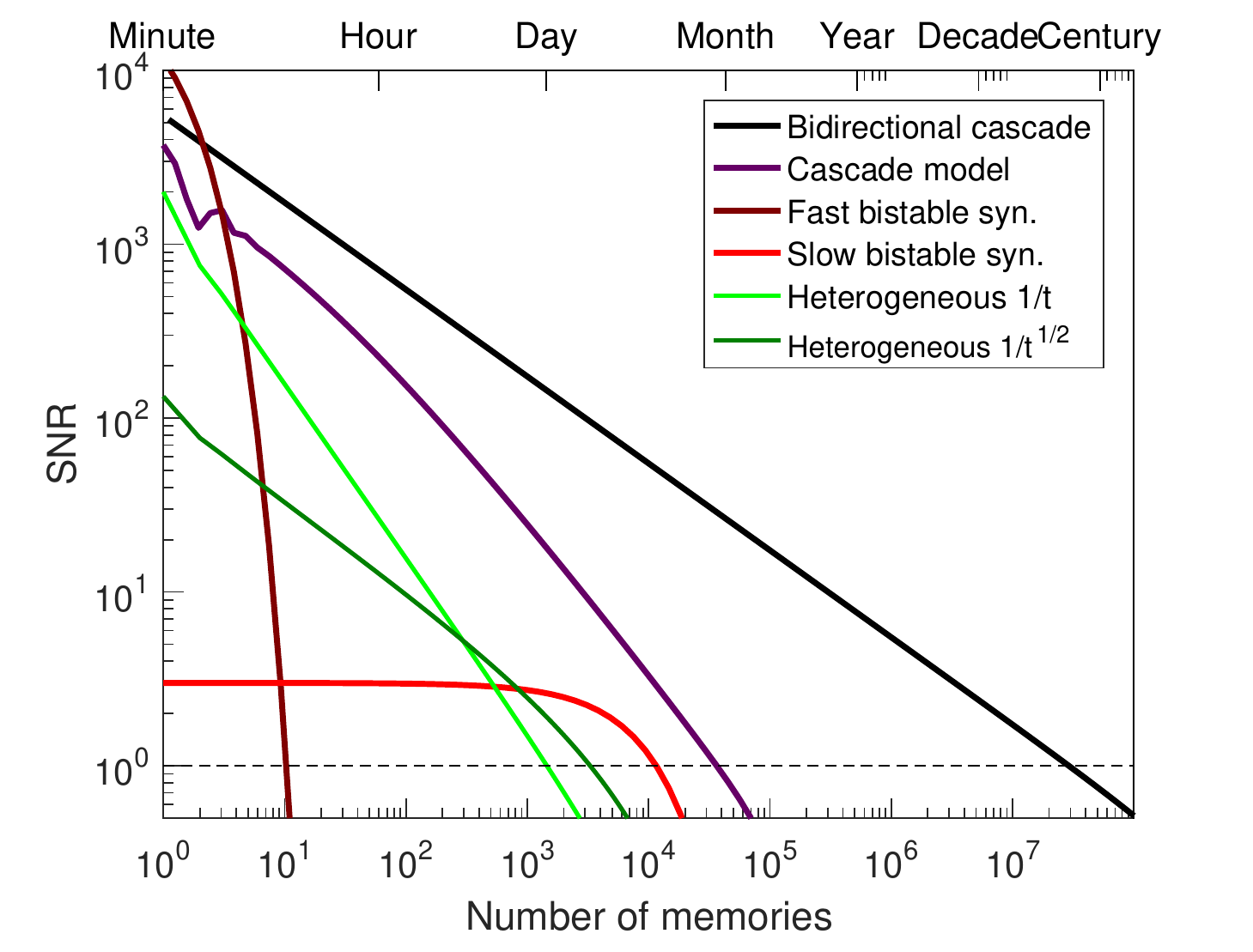}}
	\caption[]{Memory trace (signal to noise ratio, or SNR) as a function of time (i.e. the age of the tracked memory) for four different models. The dashed line is an arbitrary threshold for memory retrieval: memories are basically forgotten when SNR drops below this threshold. Dark red line: fast simple bistable synapses (all synapses have $q$ near 1). The initial memory trace is large, but the decay is rapid. Light red line: slow simple bistable synapses (all synapses have small $q \sim 1/\sqrt{N}$): long memory lifetime but small initial memory trace. Purple: cascade model, with a large initial memory trace, power law decay ($1/t$), and long memory lifetimes. Black: bidirectional cascade model: power law decay ($1/\sqrt{t}$) and the initial memory trace is as large as for the cascade model. 
	Light and dark green: heterogeneous population of simple bistable synapses. Light green: synapses are divided in 20 equal size subpopulations, each characterized by a different value of the learning rate $q$ ($q=0.6^{(k-1)}, k=1,...,20$). The SNR decays as $1/t$ and the scaling properties are the same as for the cascade model. Dark green: same number of subpopulations, but now their size increases as $q$ becomes smaller (the size is proportional to $1/\sqrt{q}$). The decay is slower ($1/\sqrt{t}$) compared to the heterogeneous model with equal size subpopulations, however the initial SNR is strongly reduced as it scales as $N^{1/4}$. For this model the memory lifetime scales as $\sqrt{N}$. To give an idea of the timescales that might be involved, for all curves we assumed that new uncorrelated memories are stored at the arbitrary rate of one every minute.
}\label{snr}
\end{figure}

\subsection{The plasticity-stability tradeoff} 

In discussing the capacity limitations of any memory model, it is important to appreciate a tradeoff between two desirable properties, plasticity and stability \cite{cg91}. To reflect this tradeoff, we characterize memory performance by two quantities \cite{fda05,fa07,bf16,arbib16}. One is the strength of the memory trace right after a memory is stored. This quantity reflects the degree of plasticity in the system, that is its ability to store new memories. The other quantity is memory lifetime, which reflects the stability of the system for storing memories over long times.

To better understand the trade-off it is important to define more precisely what we mean by memory strength.
In the standard memory benchmark, the strength of a particular memory trace is estimated in a particular situation in which memories are assumed to be random and uncorrelated. One of the reasons behind this assumption is that it allowed theorists to perform analytic calculations. However, it is a reasonable assumption even when more complex memories are considered. Indeed, storage of new memories is likely to exploit similarities with previously stored information (consider e.g.~semantic memories). Hence the information contained in a memory is likely to be preprocessed, so that only its components that are not correlated with previously stored memories are actually stored. In other words, it is more efficient to store only the information that is not already present in our memory. As a consequence, it is not unreasonable to consider memories that are unstructured (random) and do not have any correlations with previously stored information (uncorrelated). 

\subsubsection{Memory traces: signal and noise} Consider now an ensemble of $N$ synapses which is exposed to an ongoing stream of random and uncorrelated modifications, each leading to the storage of a new memory defined by the pattern of $N$ synaptic modifications potentially induced by it.  One can then select arbitrarily one of these memories and track it over time. The selected memory is not different or special in any way, so that the results for this particular memory apply equally to all the memories being stored.

To track the selected memory one can take the point of view of an ideal observer that knows the strengths of all the synapses relevant to a particular memory trace \cite{f02,fda05}. In the brain the readout is implemented by complex neural circuitry, and the strength of the memory trace based on the ideal observer approach may be significantly larger than the memory trace that is actually usable by the neural circuits. However, given the remarkable memory capacity of biological systems, it is not unreasonable to assume that the readout circuits perform almost optimally. Moreover, there are situations in which the ideal observer approach predicts the correct scaling properties of the memory capacity of simple neural circuits that actually perform memory retrieval.
 
More formally we define the memory signal of a particular memory that was stored at time $t^\mu$ as the overlap (or similarity) between the pattern of synaptic modifications $\Delta w_i$ imposed by the event and the current state of the synaptic weights $w_i$ at time $t$:
\[ \label{DefSignalMT}
{S}^{\mu}(t) \equiv {1 \over N} \Big\langle  \sum_{i=1}^N w_{i}(t)\,  \Delta w_{i}(t^\mu) \Big\rangle \ .
\]
Angle brackets indicate an average over the random uncorrelated patterns that represent the other memories and that make the trace of the tracked memory noisy. The noise is just the standard deviation of the overlap that defines the signal:
\[ \label{DefNoiseMT}
{N}^{\mu}(t) \equiv \sqrt{ {1 \over N^2}\left\langle  \Big{(} \sum_{i=1}^N w_{i}(t)\,  \Delta w_{i}(t^\mu) \Big{)}^2 \right\rangle - {S}^{\mu}(t)^2 } \ .
\]
The quantity gives the strength of the trace of memory $\mu$ will then be ${S}/{N}$, the signal to noise ratio (SNR) of a memory. 

\subsubsection{The initial signal to noise ratio: plasticity}
The initial SNR is then the SNR of a memory immediately after it has been stored, when it is most vivid. Highly plastic synapses allow for large initial SNR. Typically, in many realistic models, the SNR decreases with the memory age, so the initial SNR is often the largest SNR. It is desirable to have a large SNR (and hence a large initial SNR) because the SNR is related to the ability to retrieve a memory from a potentially noisy cue (see e.g.\cite{h82,af94}). Typically there is a threshold above which a memory becomes retrievable. This threshold depends on the architecture and the dynamics of the neural circuits that store the memory, but also on the nature of the cue that triggers memory retrieval. Highly effective cues can retrieve easily the right memory, whereas small retrieval cues might lead to the recall of the wrong memory. In the case of random uncorrelated memories it is possible to define more precisely what an effective cue is. For example, it is possible to train a perceptron to classify random input patterns and then retrieve memories by imposing on the input neurons degraded versions of the stored patterns. Degraded inputs can be obtained, for example, by changing randomly the activation state of a certain fraction of input neurons. The input patterns that are most similar to those used during training and hence stored in memory are the most effective retrieval cues. They are more likely to be classified correctly than highly degraded inputs. Higher SNR means a better ability to tolerate degradation. More quantitatively, the minimum overlap between the input and the memory to be retrieved that can be tolerated (i.e.~that produces the same response as the stored memory) is inversely proportional to the SNR\cite{Krauth1988,bf16}. This dependence demonstrates the importance of large SNRs: classifiers whose memory SNR is just above retrieval threshold can correctly recognize the inputs that have been used for training, but they will not necessarily generalize to degraded inputs. For generalization higher SNRs are needed.

\subsubsection{Memory lifetime and stability} 

Now that we have introduced a quantity that reasonably represents memory strength, we can also define more precisely the memory lifetime as the maximal time since storage over which a memory can be detected, i.e.~for which the SNR is larger than some threshold. Stable memories have long memory lifetime. The SNR threshold, as discussed above, depends on the details of the neural circuit and on the nature of the stored memories. However, the scaling properties of the memory performance do not depend on the precise value of the threshold. If new memories arrive at a constant rate, the lifetime is proportional to the memory capacity, because memories that have been stored more recently than the tracked one will have a larger SNR, and hence if the tracked memory is likely to be retrievable, so are more recent ones. The scaling of memory signal with memory age, the scaling of the initial SNR and memory lifetime with $N$ are reported for the models discussed in this article in Table \ref{comparisontable}. The actual memory capacity of neural networks will depend on many details and in particular on the neural dynamics. It is only in the recent years that investigators started to consider what is the optimal dynamics for memory retrieval \cite{Amit2010,Savin2014}.

\subsubsection{Unbounded synapses}

In the case of the models of the 80's, like the Hopfield model \cite{h82}, the memory signal is constant over time, despite the storage of new uncorrelated memories. Memories become irretrievable only because the memory noise becomes too large (the noise increases as $\sqrt{t}$) due to the interference between too many random memories. The SNR also increases with the size of the network. More specifically it is proportional to $\sqrt{N}$, where $N$ is the number of independent synapses. This means that the SNR crosses the retrieval threshold at a time $t$ that is proportional to $N$, which is a long memory lifetime if one considers that $N$ can be very large in biological brains (in the human brain the number of synapses can be of the order of $10^{15}$). This huge memory capacity is due to peculiar dependence of the memory signal on the number of stored memories: as new memories are stored, the signal always remains constant. This peculiarity comes from the assumption that the synaptic weights can grow unboundedly over time, which is clearly unrealistic for any biological system. When reasonable bounds are imposed (biological synapses are estimated to have no more than 26 distinguishable states \cite{BartolJr2015}), then the situation is very different, and the memory signal decays very rapidly with time, as discussed in the next section.

\subsubsection{Bounded synapses} 

Consider a switch-like synapse whose weight has only two values (i.e. the synapse is bistable as it can be either potentiated or depressed). This might sound like a pathological case, but it is actually representative of what happens in a large class of realistic synaptic models (see below). Suppose that a particular pattern of pre- and postsynaptic activity modifies a synapse if it is repeated over a sufficient number of trials.  The parameter $q$, which we use to characterize how labile a synapse is to change, is the probability that this pattern of activity produces a change in a synapse on any single trial. Because synapses with large $q$ values change rapidly, we call them fast, and likewise synapses with small $q$ are termed slow.  This maps a range of $q$ values to a range of synaptic timescales. For a population of synapses with a particular value of $q$, the strength of the memory trace (i.e. the SNR) at the time of storage is proportional to $q$. The memory signal decays exponentially with time, with a time constant that is proportional to $1/q$. Hence the memory lifetime goes as $1/q$. This inverse dependence is a mathematical indication of the plasticity-stability tradeoff. In non-mathematical terms, synapses that are highly labile quickly create memory traces that are vivid right after they are stored but that fade rapidly (Fig.~\ref{snr} - fast bistable synapses). Synapses that resist change and are therefore slow are good at retaining old memories, but bad at representing new ones (Fig.~\ref{snr} - slow bistable synapses).  In Fig.~\ref{snr} we plotted the memory SNR in these two cases. Notice that the horizontal and vertical scales in the figure are both logarithmic so all the differences seen are large. For example, fast synapses have an initial memory strength that is orders of magnitude larger than slow synapses. For fast synapses it is proportional to $\sqrt{N}$, where $N$ is the number of independent synapses, whereas for slow synapses it does not scale at all with $N$. However, the memory lifetime is orders of magnitude smaller for fast synapses (it scales as $\log N$, compared to the $\sqrt{N}$ scaling of slow synapses). Here we discussed the case of bistable synapses, but the plasticity stability trade-off is very general and it basically applies to any reasonably realistic synaptic model. For example, for synapses that have to traverse $m$ states before they reach the bounds, the memory capacity increases at most by a factor $m^2$, but it is still logarithmic in $N$ \cite{fa07}. The logarithmic dependence is preserved also when soft bounds are considered \cite{fa07} (see also \cite{van2012soft} for an interesting comparison between hard and soft bound synapses).  Given the generality of the plasticity-stability trade-off, how can we rapidly memorize so many details about new experiences and then remember them for years?

\begin{table}[hp]
	\vspace{2.5cm}
	\begin{center}
		\begin{tabular}{| l | c | c | c | c | c |}
			\hline
			& {\bf Time dep. } & {\bf Initial}   & {\bf Memory } & {\bf Max.\# of} &{\bf \# of} \\
			& {\bf of $\mathcal{S}$} & $\mathcal{S/N}$ & {\bf lifetime} & {\bf states} & {\bf vars}\\
			&                  &                       &                & {\bf per var.} & \\ \hline
			
			Unbounded            & const.                & n/a   & $N$ & $N$ & 1\\  \hline
			Large bounds         & $e^{-t/\tau}$                & $\leq\sqrt{N}$  & $N$ & $\sqrt{N}$ & 1\\ \hline
			Bistable (fast)     & $e^{-q t}$   & $\sqrt{N}$ & $\log(N)$ & 2 & 1\\  \hline
			Bistable (slow)     & $e^{-q t}$   & $\mathcal{O}(1)$ & $\sqrt{N}$ & 2 & 1\\  \hline	
			Heterogeneous I       & $1/t$      & ${\sqrt{N} \over \log{N}}$ & ${\sqrt{N} \over \log{N}}$ & 2 & 1 \\ \hline	
			Heterogeneous II     & $1/\sqrt{t}$      & $N^{1/4}$ & $\sqrt{N}$ & 2 & 1 \\ \hline						
			Multistage           & $1/t$      & $\sqrt{{N} \over \log{N}}$ & $\sqrt{{N} \over \log{N}}$ & 2 & 1\\ \hline	
			Multistate (hard bounds)  & $e^{-t/m^2}$ & $\sqrt{N}/m$ & $m^2 \log N$ & $m$ & 1 \\ \hline			
			Multistate (soft bounds)  & $e^{-t/m}$ & $\sqrt{N/m}$ & $m \log N$ & $\sim m$ & 1 \\ \hline			
			
			Cascade model        & $1/t$      & ${\sqrt{N} \over \log{N}}$ & ${\sqrt{N} \over \log{N}}$ & $\log(N)$ & 1\\ \hline
			Bidirectional cascade model      & $1/\sqrt{t}$ & $\sqrt{{N} \over \log{N}}$ & ${N \over \log{N}}$ & $\sqrt{\log(N)}$  & $\log(N)$ \\  \hline
		\end{tabular}
	\end{center}
	\caption{{\small Approximate scaling properties of different synaptic models. $\mathcal S$ is the memory signal, the initial $\mathcal S/N$ is the memory strength immediately after a memory is stored, and memory lifetime is defined as the time at which the SNR goes below the memory retrieval threshold. The `Unbounded' refers to models in which the synaptic variables can vary in an unlimited range, as in the Hopfield model \cite{h82}. In the case of the Hopfield model, there is no steady state, so the initial signal to noise ratio is large (as given in the table) really only for the first few memories. As more memories are stored, the noise increases, and the SNR decreases as $1/\sqrt{t}$, where $t$ is the total number of stored memories. The large bound case refers to the case in which the dynamical range of each synapse is at least of order $\sqrt{N}$ \cite{p86}. $\tau$ is of the order of $N$, and hence very large. Bistable synapses have two stable synaptic values and the transitions between them are stochastic \cite{t90,af92,af94,of13}. Fast synapses exhibit a large learning rate $q$ (i.e.~a transition probability of $\mathcal{O}(1)$), whereas slow synapses are characterized by the slowest possible learning rate (i.e.~the smallest transition probability that keeps the initial signal to noise ratio above threshold, which is $q=\mathcal{O}(1/\sqrt{N})$). In the heterogeneous models I and II\cite{fda05,rf13,bf16} the synapses have different learning rates, see Figure \ref{snr} for more details.  The multistage model is a heterogenous model in which the information about memories is progressively transferred from fast to slow synapses\cite{rf13}. The multistate models are described in \cite{fa07}. The cascade model is described in \cite{fda05} and the bidirectional cascade model in \cite{bf16} (see also the main text). Although the approximate scaling of the heterogeneous model is the same as for the cascade, the latter performs significantly better \cite{fda05}. It is important to remember that two models with the same scaling behavior may not work equally well, as the coefficients in front of the factors reported in the table might be quite different. However, it is unlikely that a model with a better scaling behavior would perform worse, as $N$ is assumed to be very large.}}
	\label{comparisontable}
\end{table}

\subsection{Cascade model of synaptic plasticity: the importance of the complexity of synaptic dynamics}

The solution proposed in \cite{fda05} is based on the idea that if we want the desirable features of both the fast and the slow synapses, we need synaptic dynamics that operates on both fast and slow timescales. Inspired by the range of molecular and cellular mechanism operating at the synaptic level, in the model proposed in \cite{fda05}, called the "cascade model",  $q$ depended on the history of synaptic modifications.  Although all the synapses in this model are described by the same equations, at any given time their properties are heterogeneous because their different histories give them different values of $q$ (metaplasticity). This improves the performance of the model dramatically and it suggests why synaptic plasticity is such a complex and multi-faceted phenomenon. The cascade model is characterized by a memory signal that decays as $1/t$. Both the initial SNR and the maximum memory lifetime scale as $\sqrt{N}$, where $N$ is the number of synapses.

The cascade model is an example of a complex synapse that does significantly better than simple synapses. However, its scaling properties are not different from those of a heterogeneous population of simple synapses in which different synapses are characterized by different values of $q$ \cite{fda05,rf13} (Fig \ref{snr}, see heterogenous models with $1/t$ decay). The interactions between fast and slow components increase significantly the numerical value of the SNR, but not its scaling properties. It is only with the recent bidirectional cascade model described below that one can improve scalability.

\subsection{The bidirectional cascade model of synaptic plasticity: complexity is even more important} 

\begin{figure}
	\centerline{\includegraphics[width=4.5in]{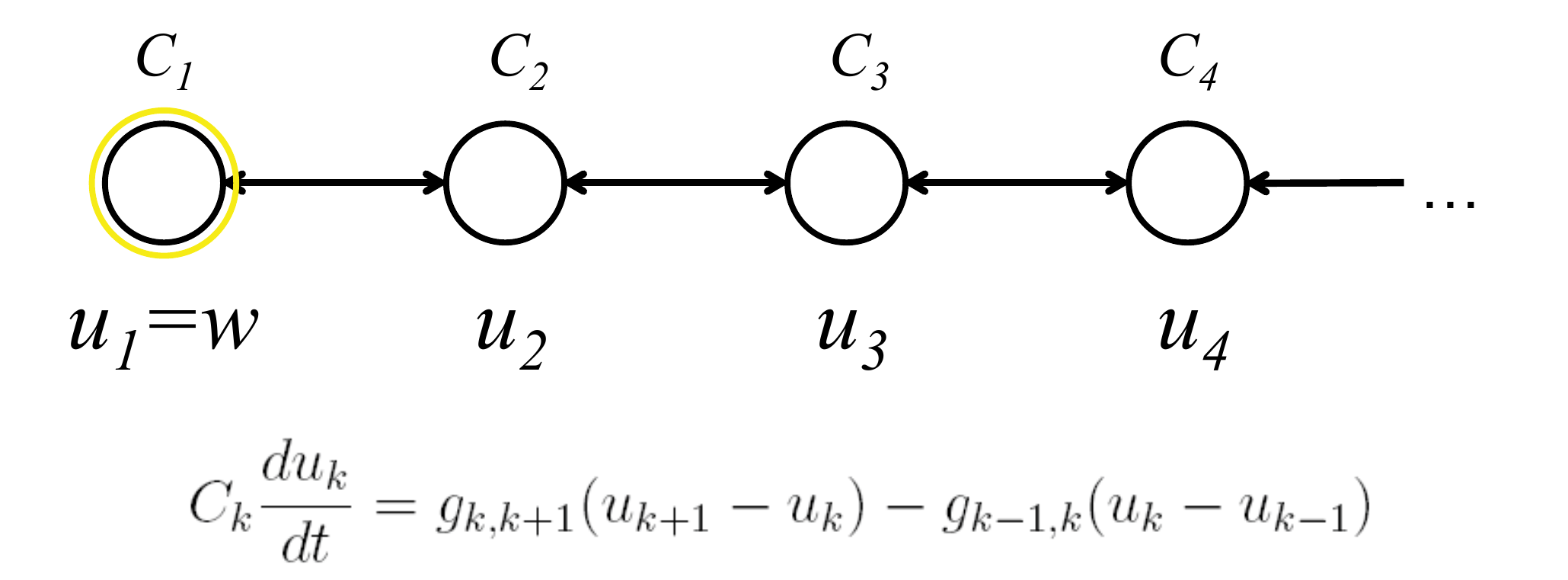}}
	\caption[]{\small The bidirectional cascade model: The dynamical variables $u_k$ represent different biochemical processes that are responsible for memory consolidation ($k=1,...,m$, where $m$ is the total number of processes). They are arranged in a linear chain and interact only with their two nearest neighbors (see differential equation), except for the first and the last variable. The first one interacts only with the second one (and is also coupled to the input), while the last one interacts only with the penultimate one. Moreover, the last variable $u_m$ has a leakage term that is proportional to its value (obtained by setting $u_{m+1}=0$). The parameters $g_{k,k+1}$ are the strengths of the bidirectional interactions (double arrows). Together with the parameters $C_k$ they determine the timescales on which each process operates. The first variable $u_1$ represents the strength of the synaptic weight.
	}
\label{bf}
\end{figure}

Bidirectional cascade models are actually a class of functionally equivalent models that are described in \cite{bf16}. Fig. \ref{bf} shows one possible implementation, a simple chain model that is characterized by multiple dynamical variables, each representing a different biochemical process. The first variable, which is the most plastic one, represents the strength of the synaptic weight. It is rapidly modified every time the conditions for synaptic potentiation or depression are met. For example, in the case of STDP, the synapse is potentiated when there is a presynaptic spike that precedes a postsynaptic action potential. The other dynamical variables are hidden (i.e. not directly coupled to neural activity) and represent other biochemical processes that are affected by changes in the first variable. In the simplest configuration, these variables are arranged in a linear chain, and each variable interacts with its two nearest neighbors. These hidden variables tend to equilibrate around the weighted average of the neighboring variables. When the first variable is modified, the second variable tends to follow it. In this way a potentiation/depression is propagated downstream, through the chain of all variables. Importantly, the downstream variables also affect the upstream variables as the interactions are bidirectional. The dynamics of different variables are characterized by different timescales, which are determined in the simple example of Fig. \ref{bf} by the $g$ and $C$ parameters. More specifically, the variables at the left end of the chain are the fastest, and the others are progressively slower. When the parameters are properly tuned, the initial SNR scales as $\sqrt{N}$, as in the cascade model previously discussed, but the memory lifetime scales as $N$, which, in a large neural system, is a huge improvement over the $\sqrt{N}$ scaling of previous models. The memory decay is approximately $1/\sqrt{t}$, as shown in Fig. \ref{snr}. The model requires a number of dynamical variables that grows only logarithmically with $N$ and it is robust to discretization and to many forms of parameter perturbations. The model is significantly less robust to biases in the input statistics. When the synaptic modifications are imbalanced the decay remains almost unaltered, but the SNR curves are shifted downwards. The memory system is clearly sensitive to imbalances in the effective rates of potentiation and depression.

In the bidirectional cascade model the interactions between fast and slow variables are significantly more important than in previous models. Indeed, it is possible to build a system with non-interacting variables that exhibits a $1/\sqrt{t}$ decay. However, this requires disproportionately large populations of slow variables, which greatly reduce the initial SNR, which would scale only as $N^{1/4}$. This leads to memory lifetimes that scale only like $\sqrt{N}$.

\subsection{Biological interpretations of computational models of complex synapses}

One possible interpretation of the dynamical variables $u_k$ is that they represent the deviations from equilibrium of chemical concentrations. The timescales on which these variables change would then be determined by the equilibrium rates (and concentrations) of reversible chemical reactions. However, for the slowest variables, which vary on timescales of the order of years, it is probably necessary to consider biological implementations in which the $u_k$ correspond to multistable processes. For example, the slowest variable could be discretized, sometimes with only two levels \cite{bf16}, and hence they could be implemented by a bistable process, which would allow for very long timescales \cite{crick1984,miller05}. For a small number of levels that is larger than two, one could combine multiple bistable processes or use slightly more complicated mechanisms \cite{Shouval2005}. These biochemical processes could be localized in individual synapses, and recent phenomenological models indicate that at least three such variables are needed to describe experimental findings \cite{Ziegler2015}.

However, these processes could also be distributed across different neurons in the same local circuit or even across multiple brain areas. The interaction between two coupled $u_k$ variables could be mediated by neuronal activity, such as the widely observed replay activity (see e.g.\cite{rf13}). In the case of different brain areas, the synapses containing the fastest variables might be in the medial temporal lobe, e.g.~in the hippocampus, and the synapses with the slowest variables could reside in the long-range connections in the cortex.

Several experimental studies on long term synaptic modifications have revealed that synaptic consolidation is not a unitary phenomenon, but consists of multiple phases. One particularly relevant example is related to studies on hippocampal plasticity and more specifically to what is known as the synaptic tagging and capture (STC) hypothesis, which explains several experimental observations. According to the STC hypothesis, LTP consists of at least four steps \cite{Reymann2007, Redondo2011}: first, the expression of synaptic potentiation with the setting of a local synaptic tag; second, the synthesis and distribution of plasticity related proteins (PrPs); third, the capture of these proteins by tagged synapses; and forth, the final stabilization of synaptic strength. Phenomenological models \cite{clopath08, Barrett2009, Ziegler2015} of STC comprise all four steps, and can explain experiments on the induction of protein synthesis dependent late LTP. The model dynamics of \cite{clopath08,Barrett2009} are characterized by four dynamical variables: the first two are tag variables, one for LTP and one for LTD. They could correspond to two variables that are modified to induce LTP and LTD. The authors of \cite{clopath08} hypothesized that a candidate molecule involved in the tag signaling could be CaMKII. The third variable describes the process that triggers the synthesis of PrPs and the fourth one the stabilization of the synaptic modification. A candidate protein involved in the maintenance of potentiated hippocampal synapses is the protein kinase M$\zeta$ (PKM$\zeta$). The PrPs that are known to be implicated in learning and plasticity include at least activity regulated cytoskeleton-associated protein (ArC), Homer1a and the AMPAr ($\alpha$-amino-3-hydroxyl-5-methyl-4-isoxazole-propionate receptor) subunit Glur1 \cite{Redondo2011}. This means that the variables of these phenomenological models should not be interpreted as concentrations of single molecules, but should be viewed as ``reporters" indicating important changes in the molecular configuration of the synapse (see the Discussion of \cite{Ziegler2015}). 

\subsection{Optimality}

The approximate $1/\sqrt{t}$ decay of the memory trace exhibited by the model in \cite{bf16} is the slowest allowed among power-law decays. Slower decays lead to synaptic efficacies that accumulate changes too rapidly and grow without bound. Interestingly, one can prove (see Suppl. Info. of \cite{bf16}) that the $1/\sqrt{t}$ decay maximizes the area between the log-log plot of the SNR and the threshold for memory retrieval (Fig. \ref{snr}).

This statement is true not only when one restricts the analysis to power laws, but also when all possible decay functions are considered. The rationale for maximizing the area under the log-log plot of the SNR can be summarized as follows: while we want to have a large SNR to be able to retrieve a memory from a small cue (see \cite{Krauth1988,bf16} and the discussion above about the importance of large initial SNR), we do not want to spend all our resources making an already large SNR even larger. Thus we discount very large values by taking a logarithm. Similarly, while we want to achieve long memory lifetimes, we do not focus exclusively on this at the expense of severely diminishing the SNR, and therefore we also discount very long memory lifetimes by taking a logarithm. While putting less emphasis on extremely large signal to noise ratios and extremely long memory lifetimes is very plausible, the use of the logarithm as a discounting function is of course arbitrary. It is interesting to consider also the case in which the SNR is not discounted logarithmically, i.e. when one wants to maximize the area under the log-linear plot of the SNR. In this situation, the optimal decay is faster, namely $1/t$, as in some synaptic models \cite{rf13,fda05}.

\subsection{Best realistic models}

As discussed above, some of the synaptic models studied in the 80's exhibited a huge memory capacity because of the unrealistic assumption that the synaptic weights could vary in an unlimited range. For any reasonably realistic model all the dynamic variables should vary in a limited range and they cannot be modified with arbitrary precision. In \cite{lg13} the authors considered a very broad class of realistic models with binary synaptic weights and multiple discrete internal states. They used an elegant approach to derive an envelope for the SNR that no realistic model can exceed. More specifically, they considered synaptic dynamics that can be described as a Markov chain. They assumed that the number of states $M$ of this Markov chain is finite, as required for any realistic model. The envelope they derived starts at an initial SNR of order $\sqrt{N}$, where $N$ is the number of independent synapses, and from there slowly decays as an exponential $\sim \exp(-t/M)$ 
up to a number of memories of order $M$, after which it decays as a power law $\sim t^{-1}$. The envelope was derived by determining the maximal SNR for every particular memory age. Hence it is not guaranteed that there exists a model that has this envelope as its SNR curve. The memory lifetime of a Markov chain model with $M$ internal states cannot exceed $\mathcal{O}(\sqrt{N} M)$.

These results indicate that one possible way to achieve a large SNR is to take advantage of biological complexity as in the bidirectional cascade model \cite{bf16}. Indeed when these models are discretized and described as Markov chains, the number of states $M$ can grow exponentially with the number of dynamical variables. Large $M$s can then be achieved even when each individual variable has a relatively small number of states (i.e. a realistically low precision). In the case of the bidirectional cascade model the number of variables and the number of states of each variable are required to grow with $N$, but very slowly (the number of variables should scale as $\log N$ and the number of states per variable scales at most as $\sqrt{\log N}$).

\subsection{The role of sparseness}

The estimates discussed in the previous sections are based on the assumption that the patterns of desirable synaptic modifications induced by stimulation are dense and most synapses are affected. This could be a reasonable assumption when relatively small neural circuits are considered, but in large networks it is likely that only a small fraction of the synapses are significantly modified to store a new memory. Sparse patterns of synaptic modifications can strongly reduce the interference between different memories, and hence lead to extended memory lifetimes. It is interesting to consider the case of random uncorrelated memories whose neural representations are sparse,  i.e. with a
small fraction $f$ of active neurons \cite{w69,tf88,Treves1990,tr91,af94,bcf98,am02,bf07,lk06,lk08,Hawkins09,dab14,bf16}. For many reasonable learning rules, these neural representations imply that the pattern of synaptic modifications is also sparse (e.g. if the synapses connecting two active neurons are potentiated, then only a fraction $f^2$ of the synapses is modified). There are also situations in which sparseness can be achieved at the dendritic level \cite{wm09} and it does not require sparseness at the neural level.

In all these cases the memory lifetime can scale almost quadratically with the number $N_{n}$ of neurons when the representations are sparse enough (i.e.~when $f$, the average fraction of active neurons, scales approximately as $1/N_{n}$). This is a significant improvement over the linear scaling obtained for dense representations. However, this capacity increase entails a reduction in the amount of information stored per memory and in the initial SNR. Scaling properties of different models are summarized in Table \ref{t2}.

The beneficial effects of sparseness that led to this improvement in memory performance are at least threefold: the first one is a reduction in the noise, which occurs under the assumption that during retrieval the pattern of activity imposed on the network reads out only the $f\,N_n$ synapses (selected by the $f\,N_n$ active neurons) that were potentially modified during the storage of the memory to be retrieved. The second one is the sparsification of the synaptic modifications, as for some learning rules it is possible to greatly reduce the number of synapses that are modified by the storage of each memory (the average fraction of modified synapses could be as low as $f^2$). This sparsification is almost equivalent to changing the learning rate, or to rescaling forgetting times by a factor of $1/f^2$. The third one is a reduction in the correlations between different synapses. This third benefit can be extremely important given that in many situations the synapses are correlated even when the neural patterns representing the memories are uncorrelated (e.g. the synapses on the same dendritic tree could be correlated simply because they share the same post-synaptic neuron\cite{af94,Savin2014}). These correlations can be highly disruptive and can compromise the favorable scaling properties discussed above.

It is important to remember that $f$ has to scale with the number of neurons of the circuit in order to achieve a superlinear scaling of the capacity. While $f \sim 1/N_n$ may be a reasonable assumption which is compatible with electrophysiological data when $N_n$ is the number of neurons of the local circuit, this is no longer true when we consider neural circuits of a significantly larger size. Moreover, sparseness can also be beneficial in terms of generalization (see e.g.\cite{olshausen2004sparse}), but only if $f$ is not too small \cite{brf13}. For these reasons, sparse representations are unlikely to be the sole solution to the memory problem. Nevertheless, plausible levels of sparsity can certainly increase the number of memories that can be stored, and this advantage can be combined with those of synaptic complexity.

Sparseness is typically assumed to be a property of the random uncorrelated neural representations that are considered for the estimates of memory capacity. However, it might also be the result of a pre-processing procedure that extracts a sparse uncorrelated component of memories which have a dense representation. In our everyday experiences, most of the new memories are similar to previously stored ones. This is the typical situation in the case of semantic memories, which contain information about categorical and functional relationships between familiar objects. For this type of memories we can utilize our previous knowledge about the objects so that we can store only the information about the relations between them (see e.g.\cite{mmo95}). In other words, we can clearly take advantage of the correlations between the new memory and the previously stored ones that encode the relevant objects. An efficient way of storing these memories is to exploit all possible correlations of this type, and then store only the memory component whose information is incompressible. This component, containing less information than the whole memory, can probably be represented with a significantly sparser neural representation. Memories are probably actively and passively reorganized to separate the correlated and the sparse incompressible part of the storable information. Modeling this process of reorganization is of fundamental importance and it has been subject of several theoretical studies \cite{mmo95,oreillyfrank2006,Kali2004,Battaglia2011a}). 

\begin{table}[hp]
	\vspace{2.5cm}
	\begin{center}
		\begin{tabular}{| l | c | c | c | c | c | }
			\hline
									& {\bf Time dep. }       & {\bf Initial}   & {\bf Min.} & {\bf Memory } & {\bf Tot.} \\
									& {\bf of $\mathcal{S}$} & $\mathcal{S/N}$ & $f$        &{\bf lifetime} & {\bf info.}\\ \hline
			
			Unbounded               & const.           & n/a        & $1/N$        & $N^2$ & $N$ \\  \hline
			Bistable $\pm 1$ & $e^{-t/f^2}$     & $f\sqrt{N}$ & $1/\sqrt{N}$ & $N$   & $\sqrt{N}$ \\  \hline
			Bistable $0,1$   & $e^{-t/f^2}$     & $\sqrt{Nf}$ & $1/N$ & $N^2$ & $N$\\  \hline
			Cascade model $0,1$          & $1/( t f^2)$      & ${\sqrt{Nf}}$ & $1/N$ & $N^2$ & $N$ \\ \hline
			Bidirectional cascade model           & $1/\sqrt{t}$ & $\sqrt{{N} \over f}$ &$1/N$& $N^2$ & $N$   \\  \hline

		\end{tabular}
	\caption{{\small Approximate scaling properties of different synaptic models in the case of sparse neural representations ($f$ is the average fraction of active neurons). In addition to the quantities described in Table \ref{comparisontable}, the last column describes the total amount of information that is storable (the information per memory scales as $fN$). Min. $f$ indicates what is the smallest $f$ that allows for an initial SNR that is larger than 1. The memory lifetime and the total storable information are computed for the minimal $f$. Unbounded refers to model proposed in \cite{tf88} in which the synaptic variables can vary in an unlimited range. As in the case of the Hopfield model, there is no steady state, so we do not report an initial SNR. Bistable synapses have two stable synaptic values and the transitions between them are stochastic \cite{af94}. Synapses are fast for potentiation (the transition probability is order 1) and relatively slow for depression (the transition probability scales as $f$). The cascade model is described in \cite{bf07} for the sparse case. The bidirectional cascade model in \cite{bf16}.}}
	\label{t2}
	\end{center}
\end{table}

\pagebreak

\section{Conclusions}

Memory is a complex phenomenon and synaptic plasticity is only one of the numerous mechanisms that the brain employs to store memories and learn. However, even when one considers only synaptic plasticity, it is now clear that it involves highly diverse interacting processes that operate on a multitude of temporal and spatial scales. Currently, there are only a few models that explain how these processes are integrated to allow the nervous system to take full advantage of the diversity of its components. All these models predict that the synaptic dynamics depends on a number of variables that can be as large as the number of biochemical processes that are directly or indirectly involved in memory consolidation. In particular, the theory shows that history dependence, which is a natural consequence of the complex network of interactions between biochemical processes, is a component of synaptic dynamics that is fundamentally important for storing memories efficiently. This greatly complicates both the theoretical and the experimental studies on synaptic plasticity because the same long term change induction protocol might lead to completely different outcomes in different experiments. A low dimensional phenomenological model that describes faithfully a series of experiments might fail in describing important observations in a different situation. For this reason, we need a new approach to the study of synaptic plasticity, in which we try to consider situations in the induction protocols imitate as much as possible the long and complex series of modifications that are caused by the storage of real world memories. Theoretical models that are based on computational principles can greatly help to design and analyze these new experiments. This is probably one of the main challenges of the next years.

\section{{Acknowledgements}}

I am very grateful to M. Benna for many fundamental discussions, comments and corrections that greatly improved the quality of the article. SF is supported by the Gatsby Charitable Foundation, the Simons Foundation, the Schwartz foundation, the Kavli foundation, the Grossman Foundation and the Neuronex NSF program.

\bibliography{oxfordarticle}

\end{document}